\UseRawInputEncoding

\documentclass[%
reprint,
superscriptaddress,
amsmath,
amssymb,
aps,
]{revtex4-2}

\usepackage{graphicx}
\usepackage{dcolumn}
\usepackage{bm}
\usepackage{physics}
\usepackage{siunitx}
\usepackage{romannum}
\usepackage{amssymb}
\usepackage{hyperref}
\hypersetup{colorlinks= true,linkcolor=blue}
\begin{document}

\title{Dipole coupling of a bilayer graphene quantum dot to a high-impedance\\ microwave resonator}

\author{Max J. Ruckriegel}
\author{Lisa M. G\"achter}
\author{David Kealhofer}
\affiliation{Laboratory for Solid State Physics, ETH Z\"urich, CH-8093 Z\"urich, Switzerland}
\author{Mohsen Bahrami Panah}
\affiliation{Laboratory for Solid State Physics, ETH Z\"urich, CH-8093 Z\"urich, Switzerland}
\affiliation{Quantum Center, ETH Z\"urich, CH-8093 Z\"urich, Switzerland}
\author{Chuyao Tong}
\author{Christoph Adam}
\author{Michele Masseroni}
\author{Hadrien Duprez}
\author{Rebekka Garreis}
\affiliation{Laboratory for Solid State Physics, ETH Z\"urich, CH-8093 Z\"urich, Switzerland}
\author{Kenji Watanabe}
\affiliation{Research Center for Electronic and Optical Materials, National Institute for Materials Science, 1-1 Namiki, Tsukuba 305-0044, Japan}
\author{Takashi Taniguchi}
\affiliation{Research Center for Materials Nanoarchitectonics, National Institute for Materials Science,  1-1 Namiki, Tsukuba 305-0044, Japan}
\author{Andreas Wallraff}
\affiliation{Laboratory for Solid State Physics, ETH Z\"urich, CH-8093 Z\"urich, Switzerland}
\affiliation{Quantum Center, ETH Z\"urich, CH-8093 Z\"urich, Switzerland}
\author{Thomas Ihn}
\author{Klaus Ensslin}
\affiliation{Laboratory for Solid State Physics, ETH Z\"urich, CH-8093 Z\"urich, Switzerland}
\affiliation{Quantum Center, ETH Z\"urich, CH-8093 Z\"urich, Switzerland}
\author{Wei Wister Huang}
\affiliation{Laboratory for Solid State Physics, ETH Z\"urich, CH-8093 Z\"urich, Switzerland}

\date{\today}

\begin{abstract}
We implement circuit quantum electrodynamics (cQED) with quantum dots in bilayer graphene, a maturing material platform for semiconductor qubits that can host long-lived spin and valley states.
The presented device combines a high-impedance ($Z_\mathrm{r} \approx \SI{1}{\kohm}$) superconducting microwave resonator with a double quantum dot electrostatically defined in a graphene-based van der Waals heterostructure. 
Electric dipole coupling between the subsystems allows the resonator to sense the electric susceptibility of the double quantum dot from which we reconstruct its charge stability diagram.
We achieve sensitive and fast detection with a signal-to-noise ratio of 3.5 within \SI{1}{\us} integration time. 
The charge-photon interaction is quantified in the dispersive and resonant regimes by comparing the coupling-induced change in the resonator response to input-output theory, yielding a maximal coupling strength of $g/2\pi = \SI{49.7}{\MHz}$.
Our results introduce cQED as a probe for quantum dots in van der Waals materials and indicate a path toward coherent charge-photon coupling with bilayer graphene quantum dots.
\end{abstract}

\maketitle
%
%
%
\section{Introduction}

The canonical circuit quantum electrodynamics (cQED) system is a solid-state qubit coupled to photons in a superconducting microwave resonator~\cite{blais_circuit_2021}. 
While the large dipole moments common to many superconducting qubits enable strong coupling~\cite{wallraff_strong_2004}, even the minuscule dipole of an individual electron isolated in a semiconductor quantum dot (QD) can interact coherently with microwave photons~\cite{stockklauser_strong_2017, mi_strong_2017}.
Strong coupling enables applications such as fast charge and spin readout~\cite{zheng_rapid_2019}, high-resolution state spectroscopy~\cite{mi_high_2017}, and photon-mediated long-range interactions~\cite{harvey-collard_coherent_2022, borjans_spin_2021, dijkema_two-qubit_2023}.
Combining cQED with semiconductor QDs is therefore of practical relevance to a possible spin qubit device architecture~\cite{vandersypen_interfacing_2017}.
Hybrid cQED has been realized in semiconductor materials like GaAs~\cite{frey_dipole_2012, stockklauser_strong_2017}, silicon~\cite{samkharadze_strong_2018, yu_strong_2023}, germanium~\cite{de_palma_strong_2023, kang_coupling_2023}, and InAs~\cite{ungerer_strong_2023, ranni_dephasing_2023}.
In this work, we extend this list to van der Waals (vdW) heterostructures by demonstrating hybrid cQED experiments with QDs in bilayer graphene.

Making use of the variety of vdW materials in cQED is a promising route for quantum technologies.
For example, highly coherent quantum circuits built with hexagonal boron nitride (hBN) and $\mathrm{NbSe_2}$ have been reported~\cite{antony_miniaturizing_2021, wang_hexagonal_2022}. 
Also monolayer graphene has been integrated with microwave circuits in the form of etched nanostructures~\cite{deng_charge_2015} or gate-tunable Josephson junctions~\cite{kroll_magnetic_2018, schmidt_ballistic_2018, wang_coherent_2019}.
Here, we choose as our material bilayer graphene encapsulated with hBN, whose voltage-induced bandgap and atomically clean interfaces make it an excellent 2D material for electrostatically defined QDs~\cite{eich_spin_2018}. 
Moreover, as a host for spin qubits, bilayer graphene also promises low spin decoherence rates due to its weak spin-orbit coupling~\cite{Kurzmann2021Kondo,Banszerus2022Spin} and weak hyperfine interaction~\cite{trauzettel_spin_2007}.
Because of these properties, bilayer graphene has become a rapidly developing material platform for spin and valley qubits.
Crucial ingredients for quantum computing applications, including time-resolved charge detection~\cite{Garreis2023counting, duprez_spectroscopy_2023} and switchable Pauli spin and valley blockade~\cite{tong_pauli_2022, banszerus_particlehole_2023, tong_pauli_2023}, have been demonstrated in electrostatically defined bilayer graphene QDs. 
The spin lifetime of up to \SI{50}{\ms} in these systems~\cite{gachter_single-shot_2022} is comparable to that of semiconductor QD systems such as \Romannum{3}-\Romannum{5}~\cite{Nakajima2020,Cerfontaine2014,Nichol2017}, silicon-~\cite{xue2021computing,Zajac2018,Yoneda2018,mills2021twoqubit} and germanium-~\cite{Hendrickx2021} based heterostructures.
Bilayer graphene also offers a unique, controllable valley degree of freedom, which exhibits exceptionally long lifetimes approaching \SI{1}{\s} ~\cite{garreis_long-lived_2023}. 
The long relaxation times allow for high-fidelity spin and valley qubit readout~\cite{gachter_single-shot_2022, garreis_long-lived_2023}.
However, the detection bandwidth of the dc charge sensors has so far been limited to around \SI{2}{kHz}~\cite{Kurzmann2019}.
These characteristics make gate-based dispersive charge readout~\cite{Johmen2023Radio, banszerus_dispersive_2021} with superconducting microwave resonators attractive for bilayer graphene, where the fundamental limit to the bandwidth is the resonator linewidth.

\begin{figure*}[t]
\includegraphics{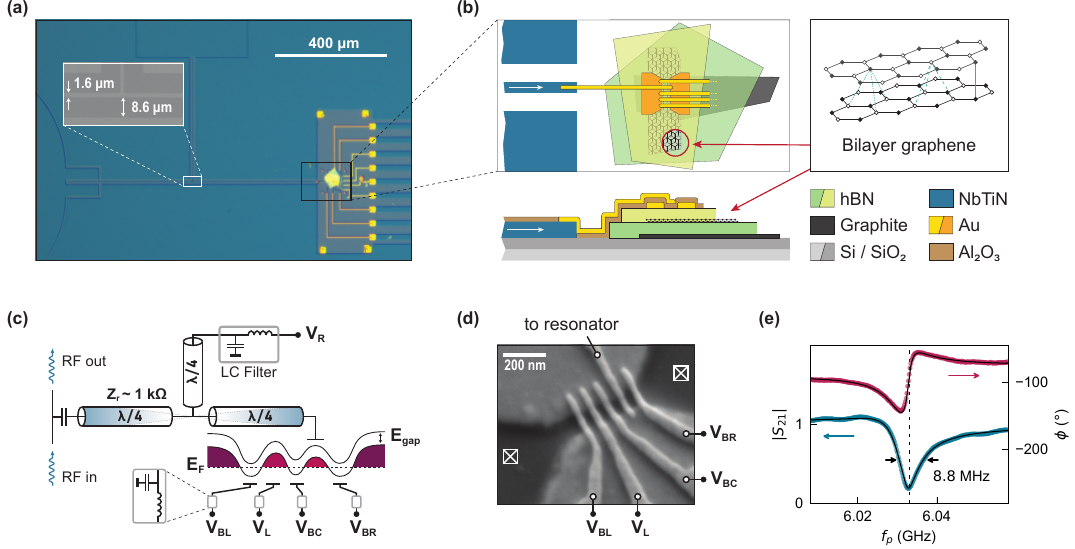}
\caption{\label{fig:1} Bilayer graphene double quantum dot (DQD) integrated with a high-impedance microwave resonator.
\textbf{(a)} Optical image of the device showing the microwave circuit fabricated from \SI{15}{\nm} thick NbTiN with the vdW material stack on the right (black rectangle).
Inset: Scanning electron micrograph (SEM) of the coplanar waveguide (CPW) at the midpoint of the half-wavelength resonator.  
\textbf{(b)} Top view and cross-section schematic of the DQD device.
The vdW material stack with bilayer graphene, encapsulated in hBN and back-gated with graphite, is deposited on a $\mathrm{Si}/\mathrm{SiO_2}$ substrate. 
Two layers of metal top gates define the DQD device.
\textbf{(c)} A simplified circuit schematic of the device in (a). 
The resonator is probed via a feedline (left).
The voltage $V_\mathrm{R}$ is applied to the plunger gate through the center conductor of the CPW via a LC-filter and quarter-wavelength tap.
The Fermi energy $E_\mathrm{F}$ of the bilayer graphene along the channel is tuned by plunger and barrier gates to form a double-well potential for holes.
\textbf{(d)} SEM of the gates on top of the vdW material stack, with voltages applied to the barrier and plunger gates labeled.
Crossed boxes indicate ohmic contacts to the bilayer graphene.
\textbf{(e)} Microwave transmission spectrum through the feedline around the resonator frequency $f_\mathrm{r}$ (dashed line). }
\end{figure*}

In this paper, we present a device that integrates an on-chip superconducting resonator with a bilayer graphene double QD (DQD). 
We use the kinetic inductance of NbTiN to realize a high-impedance resonator whose zero-point voltage fluctuations are more than four times larger than for a conventional \SI{50}{\ohm} resonator.
The larger voltage fluctuations enhance the coupling strength $g$ to the electric dipole moment of DQD state transitions. 
We demonstrate resonator-based sensing of the DQD electric susceptibility that complements transport measurements.
The high sensitivity of the measurement allows us to detect an interdot transition with a signal-to-noise ratio (SNR) of $3.5$ within \SI{1}{\us} integration time.
We estimate the minimum time $t_\mathrm{min} \approx \SI{1e-7}{\s}$ needed to achieve a SNR larger than one, four orders of magnitude faster than current dispersive gate-based sensing of QDs in bilayer graphene~\cite{banszerus_dispersive_2021} and comparable to state-of-the-art readout with on-chip resonators in silicon~\cite{zheng_rapid_2019}.
Furthermore, we measure the resonator response to the DQD dipole moment in both the dispersive and resonant cases, and extract the relevant system parameters from a least-squares fitting routine.
We achieve a bare resonator coupling strength of up to $g/2\pi = \SI{49.7}{\MHz}$ that we assume to be limited by the small lever arm difference $\beta$ of the coupling gate.
The charge decoherence rate $\gamma = \SI{643}{\MHz}$ is likely caused by high tunneling rates of electrons to the leads, effectively making this value an upper bound for $\gamma$~\cite{hecker_coherent_2023}.
This work constitutes the next technological step for bilayer graphene as a spin or valley qubit platform and marks the necessary advances to achieve coherent charge-photon coupling in vdW materials.

%
\section{Hybrid device architecture}
%
Figure \hyperref[fig:1]{1} shows the hybrid device.
We sputter \SI{15}{\nm} of NbTiN on an intrinsic silicon wafer with \SI{100}{\nm} thermal $\mathrm{SiO_2}$.
A half-wavelength resonator is patterned into the superconducting film as a coplanar waveguide with a center conductor width of \SI{1.6}{\um} and a length of \SI{900}{\um} [panel \hyperref[fig:1]{(a)}].
Based on the resonator geometry and its center frequency $f_\mathrm{r} = \SI{6.033}{\GHz}$ we estimate the resonator impedance $Z_\mathrm{r} = \sqrt{L_\mathrm{l} / C_\mathrm{l}} \approx \SI{1}{\kohm}$. 
The inductance per unit length $L_\mathrm{l}$ is dominated by the large contribution of the sheet kinetic inductance of the NbTiN film which we estimate to $L_{\square, \mathrm{kin}} \approx \SI{150}{\pico \henry}  \mathrm{/} \square$.
At one end, the resonator is capacitively coupled to a \SI{50}{\ohm} feedline. 
Close to the other end of the resonator, we fabricate the bilayer graphene DQD device, similar to previously reported samples ~\cite{eich_coupled_2018}. 

Figure~\hyperref[fig:1]{1~(b)} shows a schematic of the device. 
The vdW material stack consists of Bernal-stacked bilayer graphene, encapsulated in hexagonal boron nitride (hBN) and back-gated with graphite.
Two layers of gold gates are fabricated on top of the vdW material stack with ALD-deposited $\mathrm{Al_2O_3}$ as the intergate dielectric. 
One of the gates connects to the resonator center conductor.
We apply a voltage to this gate through a quarter-wavelength bias tap at the midpoint of the resonator. 
Electron transport through the quantum dot is probed via ohmic contacts to the graphene, achieved by etching through the top layer of hBN before metal deposition. 
To prevent microwave leakage through parasitic gate capacitances~\cite{harvey-collard_-chip_2020, holman_microwave_2020}, all gate lines are biased through on-chip low-pass filters consisting of a shunt capacitance to ground in series with the large inductance of a narrow NbTiN wire.

The sample is mounted in a dilution refrigerator that reaches the base temperature $T = \SI{10}{\milli \kelvin}$.
For resonator spectroscopy, we probe the microwave transmission through the feedline as a function of frequency $f_\mathrm{p}$.
The transmitted microwave signal undergoes a heterodyne detection scheme that allows us to measure the complex transmission amplitude  $S_{21}(f_\mathrm{p}) = A e^{i \phi} = I + i Q$.
Transmission is reduced around the resonator frequency [see Figure \hyperref[fig:1]{1 (e)}] as expected for the notch-type coupling to the feedline.
From a fit to the spectrum around $f_\mathrm{r}$~\cite{Probst2015Efficient} we extract the internal resonator decay rate $\kappa_\mathrm{int} / 2\pi = \SI{1.9}{\MHz}$ and the total linewidth $\kappa / 2\pi = \SI{8.8}{\MHz}$.
The internal quality factor $Q_\mathrm{int} = 2\pi f_\mathrm{r}/\kappa_\mathrm{int} = 3250$ is limited by the dielectric environment of $\mathrm{SiO_2}$ and $\mathrm{Al_2O_3}$~\cite{ungerer_performance_2023}.
However, we have observed that the $\mathrm{SiO_2}$ layer is necessary to improve device yield.

A negative voltage applied to the back-gate opens a bandgap in bilayer graphene and together with a pair of split-gates confines valence band holes to a narrow channel.
The second set of gates across the channel defines plunger and barrier gates.
Figure~\hyperref[fig:1]{1 (c)} shows a device schematic and Figure~\hyperref[fig:1]{1 (d)} a scanning electron micrograph of the device with the plunger and barrier gates labeled.
Each of these gates varies the local potential along the channel, placing the Fermi energy in the valence band, the bandgap, or the conduction band.
With appropriate positive voltages applied, they realize a double-well potential to host the DQD.
The plunger gate voltages $V_\mathrm{L}$ and $V_\mathrm{R}$ change the electrochemical potentials of electrons in the left and right QD.
The gate voltages $V_\mathrm{BL}$ and $V_\mathrm{BR}$ are set to place the Fermi energy in the bandgap such that these regions act as tunnel barriers between the dots and their leads. 
Similarly, the voltage at the central barrier, $V_\mathrm{BC}$, changes the coupling between the QDs. 
The gate connected to the resonator is the right QD plunger gate, biased through the resonator and low-pass filter with $V_\mathrm{R}$. 
Its capacitance to the right QD couples the DQD to the electric field fluctuations at the voltage antinode of the microwave resonator. 
This makes the resonator susceptible to the dipole moment of electrons in the QDs and thereby acts as a sensor.

\section{Dispersive sensing}
%
\begin{figure}[b]
\includegraphics{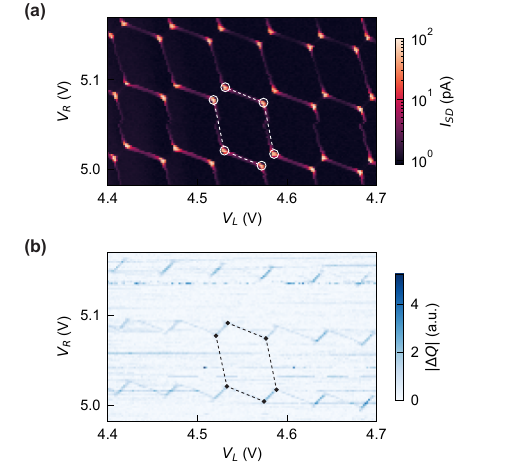}
\caption{\label{fig:2} 
DQD charge stability diagram. 
\textbf{(a)} Current $I_\mathrm{SD}$ measured through the channel at a \SI{200}{\uV} bias voltage as a function of plunger gate voltages. 
White dashed lines mark the boundaries of one hexagonal region with stable charge configuration.
Circles mark the triple points.
\textbf{(b)} Change in the quadrature component $Q$ of the microwave signal transmitted through the feedline at fixed probe frequency $f_\mathrm{p} = f_\mathrm{r}$.
The charge stability diagram is visible (marked in black) with interdot transitions connecting the triple points.}
\end{figure}
\begin{figure}[b]
\includegraphics{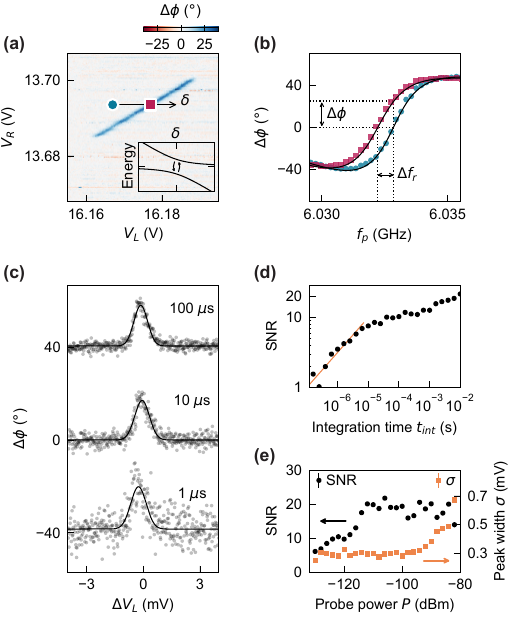}
\caption{\label{fig:3} 
Dispersive resonator shift and SNR analysis.
\textbf{(a)} Microwave response of one example interdot transition.
The black arrow indicates the detuning axis.
The inset is a schematic energy diagram of the charge qubit states as a function of QD energy detuning $\delta$.
Tunnel coupling leads to hybridization at $\delta=0$ with an energy splitting $2 t_\mathrm{c}$.
\textbf{(b)} Phase response of the resonator as a function of probe tone $f_\mathrm{p}$ at the two detunings marked by symbols in \textbf{(a)}.
Dispersive interaction with the charge qubit shifts the resonator to lower frequency at $\delta = 0$ (purple squares) compared to $\delta=\SI{4}{\mV}$ (blue circles).
\textbf{(c)} Three example traces across the transition for different integration times $t_\mathrm{int}$.
A fit of a Gaussian peak to the data (black solid lines) determines the SNR as a function of $t_\mathrm{int}$.
For $t_\mathrm{int} = \SI{1}{\us}$ the phase signal is still clearly discernible with a SNR of 3.5. 
\textbf{(d)} The SNR reduces with shorter $t_\mathrm{int}$ but stays well above unity even for integration times below \SI{1}{\us}. A linear extrapolation to SNR = 1 (orange solid line) gives $t_\mathrm{min} \approx \SI{1e-7}{\s}$.
\textbf{(e)} SNR and width $\sigma$ of the fitted Gaussian peak as a function of resonator probe power. Error bars for SNR in (d) and (e) are smaller than the marker size.}
\end{figure}
Figure \hyperref[fig:2]{2 (a)} and \hyperref[fig:2]{(b)} depict the DQD charge stability diagram measured simultaneously in dc current and resonator response.
The source-drain current $I_\mathrm{SD}$ through the device as a function of plunger gate voltages $V_\mathrm{L}$ and $V_\mathrm{R}$ shows the hexagonal pattern typical for transport through a DQD.
Within one hexagon, the number of charges in each QD is constant, and the current is suppressed due to Coulomb blockade. 
At the four boundaries of a stable charge configuration, exemplarily indicated by dashed white lines in \hyperref[fig:2]{(a)}, the electrochemical potential of one QD is resonant with the Fermi energy of the respective lead while the other dot is Coulomb blockaded, and transport can occur via co-tunneling processes through the other dot.  
High current at the triple points, marked with white circles, corresponds to the QD electrochemical potentials in both dots being aligned inside the bias window. 

Simultaneous with the current, we monitor the complex microwave transmission through the feedline with the probe frequency $f_\mathrm{p} = f_\mathrm{r}$. 
Figure~\hyperref[fig:2]{2 (b)} shows the change in the quadrature component $Q$ of the complex signal as a function of $V_\mathrm{L}$ and $V_\mathrm{R}$.
The observed pattern matches the current data with a large change in $Q$ along the lines of positive slope connecting adjacent triple points and lines of negative slope coinciding with the co-tunneling current visible in \hyperref[fig:2]{(a)}.
Along these co-tunneling lines, the resonator is susceptible to tunneling of electrons between the right QD and its lead.
On the other hand, the signal between adjacent triple points corresponds to transitions between charge configurations with the same total number of electrons.
The charge states hybridize along interdot transitions due to the tunnel coupling $t_\mathrm{c}$ between the QDs.
These hybridized states form a two-level system whose electric susceptibility changes the resonator response and therefore the transmission through the feedline at $f_\mathrm{p}$.

%
%
%
%
%
%
%
We observe this sensing signal over a wide range of gate voltages and QD configurations. 
Figure \hyperref[fig:3]{3 (a)} shows an interdot transition at higher gate voltages with a strong phase response.
In section \hyperref[chargephoton]{IV} we show that this sensing signal of the DQD electric susceptibility is in the dispersive regime.
Sweeping $V_\mathrm{L}$ across the transition (black arrow) changes the energy detuning $\delta$ between the QD electrochemical potentials in the left and right dot.
Microwave spectroscopy at two different QD detunings [Figure \hyperref[fig:3]{3 (b)}] reveals a clear shift of the resonator to lower frequencies for $\delta = 0$ (purple squares) compared to non-zero detuning (blue circles).
Measured at a fixed-frequency probe tone $f_\mathrm{p}$, the shift of the resonator frequency results in a phase change $\Delta \phi$ of more than \SI{26}{\degree}.

The sensitivity of the charge detection signal is investigated in more detail by determining the signal-to-noise ratio (SNR) as a function of integration time and probe power.
We sweep the gate voltage $V_\mathrm{L}$ across the interdot transition and reduce the measurement integration time $t_\mathrm{int}$ from \SI{100}{\ms} to \SI{100}{\ns} at a probe power of $P = \SI{-90}{dBm}$.
Figure  \hyperref[fig:3]{3 (c)} shows three examples of measured traces for different integration times.
A fit of a Gaussian function to the data (solid black lines) determines the peak height $A$ and the peak width $\sigma$. 
Subtracting the fit from the data and calculating the standard deviation gives the noise $B$ and the $\mathrm{SNR} = A/B$.
The SNR decreases with shorter $t_\mathrm{int}$ [Figure  \hyperref[fig:3]{3 (d)}] from $\mathrm{SNR} > 20$ to a value of 3.5 at $t_\mathrm{int} = \SI{1}{\us}$. 
We extrapolate the $\propto \sqrt{t_\mathrm{int}}$ trend for short $t_\mathrm{int}$ (orange solid line) and estimate the minimum integration time $t_\mathrm{min} \approx \SI{1e-7}{\s}$ needed to achieve $\mathrm{SNR} = 1$. 
This value compares well with state of the art dispersive readout of silicon QDs ~\cite{zheng_rapid_2019} and is four orders of magnitude faster than previous dispersive charge sensing in bilayer graphene~\cite{banszerus_dispersive_2021}.

Figure \hyperref[fig:3]{3 (e)} shows the dependence of the SNR on resonator probe power (black dots).
The signal strength grows with an increase in probe power up to a power of $P = \SI{-115}{dBm}$. 
Beyond this power level, the SNR remains the same.  
Additionally, when the power exceeds $P = \SI{-95}{dBm}$, thermal broadening of the peak becomes evident, marked by a significant increase in the peak width $\sigma$ (orange squares).

\section{Charge-photon coupling} 
\label{chargephoton}
%
%
%
%
%
%
%
%
%
%
%
%
%
%
Beyond charge sensing, we characterize the dipole coupling of the DQD to microwave photons in the resonator.
The gate voltage $V_\mathrm{L}$ varies the energy detuning $\delta$ between charge states of an electron being in the left or in the right QD, forming the two-level system of a canonical charge qubit with energy $E_\mathrm{q}(\delta) = \sqrt{4 t_\mathrm{c} ^2 + \delta^2}$. 
At $\delta = 0$, the states hybridize due to tunnel coupling with amplitude $t_\mathrm{c}$ and are separated by the minimal charge qubit energy $2 t_\mathrm{c}$.  
Its interaction with the resonator depends on two properties of the charge qubit: 
The coupling strength is proportional to its electric dipole moment which is maximized when the electron wavefunction is completely delocalized between the QDs at $\delta=0$.
Furthermore, the charge qubit needs to be resonant with the resonator ($E_\mathrm{q}/h = f_\mathrm{r}$) for the two systems to exchange energy, otherwise the interaction is dispersive.

\begin{figure}[b]
\includegraphics{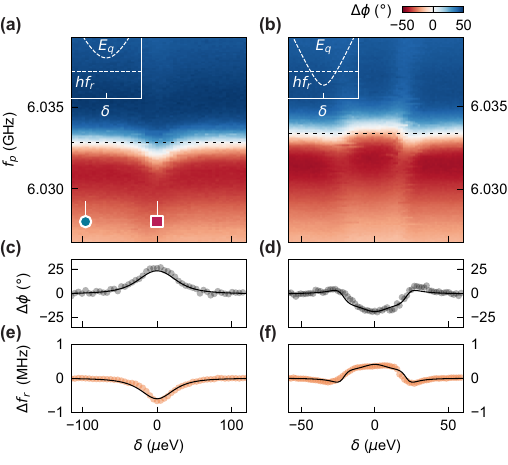}
\caption{\label{fig:4}
Comparison of charge-photon coupling for two different interdot transitions with $2t/h > f_\mathrm{r}$ (left column) and $2t/h < f_\mathrm{r}$ (right column).
\textbf{(a)} Resonator phase signal as a function of DQD detuning $\delta$ and probe frequency $f_\mathrm{p}$. 
Dispersive interaction with the charge qubit shifts the resonator frequency around $\delta = 0$.
Purple and blue marks indicate the line traces shown in Figure \hyperref[fig:3]{3 (b)}.
\textbf{(c)} Line cut of $\Delta \phi$ along detuning for fixed $f_\mathrm{p}$ indicated by the dashed black line in panel (a).  
\textbf{(e)} Extracted resonator frequency shift $\Delta f_\mathrm{r}$ as a function of detuning. 
Solid black lines overlay input-output theory calculations for best fit parameters $(g/2\pi , \gamma/2\pi, 2t/h) = (\SI{49.7}{\MHz}, \SI{643}{\GHz}, \SI{10.2}{GHz})$.   
The spectroscopy data in \textbf{(b)} shows a different interdot transition with $2t/h < f_\mathrm{r}$. 
A larger charge rearrangement was corrected in this data set. 
\textbf{(d)} Phase shift along the dashed black line in (b) and \textbf{(f)} resonator frequency shift extracted from (b). 
Calculations with best fit parameters $(g/2\pi , \gamma/2\pi, 2t/h) = (\SI{37.5}{\MHz}, \SI{1.11}{\GHz}, \SI{2.99}{GHz})$ are shown as solid black lines in (d) and (f).
}
\end{figure}

Input-output theory~~\cite{Burkard2016Dispersive} describes photons in a resonator interacting with a two-level system, both in the resonant and in the dispersive regimes. 
A comparison between experimental resonator spectroscopy data and input-output theory characterizes the hybrid system and allows us to extract its coupling and decoherence properties.
We determine some system parameters from independent measurements, such as the frequency $f_\mathrm{r}$ and the linewidth $\kappa$ of the unperturbed resonator. 
Parameters to be derived from input-output theory are the charge qubit decoherence rate $\gamma$ and the DQD tunnel coupling $t_\mathrm{c}$. 
We also leave the coupling strength $g$ as a fitting parameter that we can compare to the estimate $g/2 \pi = \frac{\sqrt{\pi}}{2} \beta f_\mathrm{r} \sqrt{\frac{Z_\mathrm{r}}{h/e^2}} \approx \SI{38}{\MHz}$. 
This value for $g$ uses the lever arm difference $\beta \approx 0.037 \pm 0.005$ of the resonator gate $V_\mathrm{R}$ to the DQD that we estimate from finite bias measurements. 

Figure \hyperref[fig:4]{4(a)} and \hyperref[fig:4]{(b)} contrast $\Delta \phi$ as a function of probe frequency $f_\mathrm{p}$ and detuning $\delta$ for two different interdot transitions that demonstrate qualitatively different behavior.
The data displayed in Figure \hyperref[fig:4]{4(a)} is from the transition in Figure \hyperref[fig:3]{3(a)}.
Purple and blue marks indicate the line traces from Figure \hyperref[fig:3]{3(b)} taken at two different detunings.
We interpret the observed resonator shift to lower frequencies around $\delta = 0$ as the effect of the charge qubit with its minimal excitation frequency $2t_\mathrm{c} / h$ being larger than the resonator frequency $f_\mathrm{r}$.
For $2t_\mathrm{c}/h \gg f_\mathrm{r}$, the interaction is dispersive and acts as a perturbation on the two subsystems.
Dipole coupling to the charge qubit shifts the resonator frequency by $2\pi \Delta f_\mathrm{r} \approx - g^2/\left(E_\mathrm{q}/\hbar - 2\pi f_\mathrm{r}\right)$. 
The black dashed line in Figure \hyperref[fig:4]{4(a)} indicates the cut for fixed probe frequency shown in panel \hyperref[fig:4]{(c)}.
From fits of the transmission spectrum at each value of $\delta$ we extract the frequency shift $\Delta f_\mathrm{r}$ as a function of detuning [panel \hyperref[fig:4]{(e)}].
The largest shift is $\Delta f_\mathrm{r} = \SI{-0.65}{\MHz}$ at $\delta=0$.
A least-squares fit of input-output theory to the full data gives best estimates of the free parameters (see Appendix \hyperref[app:fit]{C} for details). 
For the dispersive case with $2t_\mathrm{c}/h \gg f_\mathrm{r}$ the best fit is for $g/2 \pi = \SI{49.7}{\MHz}$, $\gamma/2 \pi = \SI{643}{\MHz}$ and $2 t_\mathrm{c}/ h = \SI{10.2}{\GHz}$. 
Solid lines in panels \hyperref[fig:4]{(c)} and \hyperref[fig:4]{(e)} overlay the data with $\Delta \phi$ and $\Delta f_\mathrm{r}$ calculated using this set of parameters, demonstrating reasonable agreement. 

The second data set [Figure \hyperref[fig:4]{4(b)}] was taken for a different interdot transition at different plunger gate voltages. 
The black dashed line indicates where the trace of  $\Delta \phi$ shown in panel \hyperref[fig:4]{(d)} was taken, and we again extract the resonator shift $\Delta f_\mathrm{r}$ as a function of detuning [panel \hyperref[fig:4]{(f)}].
Both $\Delta \phi$ and $\Delta f_\mathrm{r}$ showcase the qualitative differences to the interdot transition in the left column of Figure \hyperref[fig:4]{4}.
In contrast to the dispersive case, the charge qubit frequency crosses the resonator at $\delta = \pm \SI{25}{\micro eV}$ and reaches a minimal frequency $2t_\mathrm{c}/h < f_\mathrm{r}$.  
Within this range of detuning, $E_\mathrm{q}(\delta)/h$ is below the resonator frequency $f_\mathrm{r}$ so that interactions with the charge qubit shift the resonator to higher frequencies by $\Delta f_\mathrm{r} = \SI{+0.39}{\MHz}$. 
At the detuning energy where $E_\mathrm{q}(\delta)/h$ is close to $f_\mathrm{r}$, the interaction is resonant and the high decoherence rate $\gamma$ of the charge qubit reduces the resonator quality factor.  
We determine the best fitting parameters for the $2t_\mathrm{c}/h < f_\mathrm{r}$ case to be $g/2 \pi = \SI{37.5}{\MHz}$, $\gamma/2 \pi = \SI{1.11}{\GHz}$ and $2 t_\mathrm{c}/ h = \SI{2.99}{\GHz}$. 

In both the dispersive and the resonant case we obtain a value for $g/2\pi$ that is close to or even higher than our first estimate of \SI{38}{\MHz}. 
This confirms that there is no significant loss in coupling strength from impedance mismatches along the gate connected to the resonator.
Differences to the estimated coupling strength are likely introduced through the high uncertainty in estimating the lever arm difference $\beta$ from finite bias measurements.
While the gate capacitances that determine the lever arms depend on the exact tuning of the DQD, they are mostly determined by the implemented gate layout. 
Generally, we also note that the charge qubit decoherence rate $\gamma$ is large compared to the coupling strength $g$ and resonator linewidth $\kappa$, preventing the system from reaching the strong coupling regime.
We assume that the observed $\gamma$ is large because the DQD is not sufficiently isolated from the leads~\cite{Mi2017High}. 
High tunneling rates to the leads limit the average time charges spend in the DQD and act as a loss of charge qubit coherence.
Effectively, the value for $\gamma$ estimated from our measurements therefore sets an upper bound for the decoherence rate of a charge qubit in bilayer graphene.

\section{Conclusion \& Outlook}
%
We demonstrated electric dipole coupling of a bilayer graphene DQD to a high-impedance superconducting microwave resonator.
We used the on-chip resonator with low internal losses for dispersive charge sensing and observed a signal-to-noise ratio well above unity even for integration times shorter than \SI{1}{\us}.
Furthermore, we investigated the charge-photon coupling in the dispersive and resonant cases. 
By comparing spectroscopy data to input-output theory, we estimated the relevant coupling parameters.
Notably, the achieved coupling strengths are comparable to those in early-stage silicon quantum dots~\cite{mi_circuit_2017}, despite the small gate lever arms in our device.
Our analysis shows that, with $g \gg \kappa$ already satisfied, only the charge qubit decoherence $\gamma$ prevents the system from reaching the anticipated strong charge-photon coupling regime that holds the full potential of hybrid cQED. 
Better tunability of tunneling barriers and a reduction in charge decoherence should be achievable by an improved gate design with overlapping barrier and plunger gates~\cite{mi_strong_2017}.
Additionally, refining the gate layout to increase the gate lever arms would enhance the charge-photon coupling rates. 
Already with the current coupling strength, one straightforward optimization is a reduction of the resonator coupling to the feedline that dominates the total linewidth $\kappa$.
In particular, a critically coupled resonator would bring the system closer to $\kappa/2\pi \approx \Delta f$ for maximal signal visibility while maintaining a high detection bandwidth.

Our results demonstrate how hybrid cQED techniques can be used to probe bilayer graphene QDs and lead the way to coherent charge-photon coupling.
To go further, hybridization of charge and spin degrees of freedom, achieved by similar means as in silicon~\cite{mi_coherent_2018, samkharadze_strong_2018}, can enable strong spin-photon coupling also with bilayer graphene QDs.  
Looking ahead to spin and valley qubits, cQED applications such as fast state readout or long range interactions may become essential for bilayer graphene as a material platform for qubits.

\section*{Acknowledgements}
We thank P. M\"arki and T. B\"ahler as well as the FIRST staff for their technical support.
We thank Ekatarina Al-Tavil and Andr\'es Rosario Hamann for valuable discussions.
We thank A. Denisov for valuable input during writing of this manuscript.
We acknowledge financial support from the European Graphene Flagship, the ERC Synergy Grant Quantropy, and the European Union’s Horizon 2020 research and innovation program under grant agreement number 862660/QUANTUM E LEAPS and NCCR QSIT (Swiss National Science Foundation). K.W. and T.T. acknowledge support from the JSPS KAKENHI (Grant Numbers 21H05233 and 23H02052) and World Premier International Research Center Initiative (WPI), MEXT, Japan for the growth of h-BN crystals.

\appendix

\section{Device Fabrication}

The sputtering process was carried out in an Orion 8 magnetron sputtering system (AJA International, Inc.) in reactive dc mode from a NbTi (Nb/Ti 70/30 wt\%) target (ACI Alloys, Inc.) with 99.99\% purity at \SI{1.5e-8}{} mTorr base pressure. 
We initiate the process with a target cleaning step (5 min exposure to argon (Ar) plasma at 50 sccm flow rate and 100 W) and a conditioning phase (4 sccm nitrogen ($\mathrm{N_2}$) for 1 min with shutter closed) . 
The subsequent deposition stage consists of Ar/$\mathrm{N_2}$ flows at 50 sccm and 4 sccm, respectively, maintaining a power of 100 W, a pressure of 3.5 mTorr, and a working distance of 10 cm. 
The parameters were chosen to optimize the growth conditions for NbTiN films, film quality, thickness uniformity, and structural properties.

We sputter a nominally \SI{15}{\nm} thick film of NbTiN on a two-inch wafer of intrinsic silicon ($\rho > \SI{10}{\kohm \cm}$) with \SI{100}{\nm} of thermally grown $\mathrm{SiO_2}$ (Alineason Materials Technology GmbH). 
The microwave circuit is patterned using a direct-write photo-lithography system (Heidelberg Instruments DWL66+) and reactive ion etching (RIE) with $\mathrm{SF_6}/\mathrm{Ar}$.
Afterwards, we deposit gold markers and bondpads in a lift-off process and dice the wafer into $5 \times \SI{8}{\mm}$ chips. 
The vdW material stack is fabricated on the pre-patterned chips using standard mechanical exfoliation from bulk crystals and polymer-based dry transfer techniques. 
We first deposit the bottom hBN ($\sim \SI{25}{nm}$ thick) and graphite back-gate onto the chip and clean it from polymer residues.
The top hBN ($\sim \SI{35}{nm}$ thick) and bilayer graphene are placed onto the pre-deposited bottom half in a separate deposition.
Metal gates are fabricated in a lift-off process by electron beam lithography (EBL) and metal evaporation.
For ohmic contacts to the graphene, we etch hBN before metal deposition using RIE with $\mathrm{CHF_3}$.
The split-gates are made from 3/\SI{20}{\nm} Cr/Au with a \SI{100}{\nm} wide channel.
Plunger and barrier gates are also made from 3/\SI{20}{\nm} Cr/Au with \SI{25}{\nm} in width and a gate pitch of \SI{70}{\nm}. 
They are separated from the first gate layer by \SI{20}{\nm} of $\mathrm{Al_2O_3}$ grown by atomic layer deposition (ALD) at \SI{150}{\celsius}. 
The $\mathrm{Al_2O_3}$ layer is removed by wet etching to establish contact to the resonator below the dielectric.

\section{Input-Output Theory}

The complex transmission $S_{21}$ through the microwave feedline with a resonator coupled in a notch-type configuration can be derived from input-output theory~\cite{Burkard2016Dispersive} to be
$$S_{21}(f_\mathrm{p}, \delta) = 1 + \frac{i \kappa_{\mathrm{ext}}}{ 2\pi(f_\mathrm{p} - f_\mathrm{r}) - i\kappa + g^*\chi_\mathrm{e}}.$$
The resonator frequency $f_\mathrm{r}$ and total linewidth $\kappa = \kappa_{\mathrm{ext}} + \kappa_{\mathrm{int}}$ are determined from a fit to the spectrum of the unperturbed resonator~\cite{Probst2015Efficient}.
For this fit, the external coupling $\kappa_\mathrm{ext}$ is taken as a complex-valued parameter to account for an asymmetric Fano lineshape of the resonance arising from non-idealities in the resonator-feedline coupling.
The effect of the charge qubit is described by its electric susceptibility
$$\chi_\mathrm{e} = \frac{g^*}{i \gamma + 2\pi(f_\mathrm{p} - E_\mathrm{q}(\delta)/h)},$$
that considers the charge qubit decoherence rate $\gamma$ and probe detuning.
The charge qubit energy $E_\mathrm{q}(\delta) = \sqrt{4 t_\mathrm{c}^2 + \delta^2}$ depends on the interdot tunnel coupling $t_\mathrm{c}$ and the DQD energy detuning $\delta$.
We calculate $\delta$ from the lever arms $\alpha^\mathrm{L}_\mathrm{dL(R)}$ of the left plunger gate to the left (right) dot as $\delta = (\alpha^\mathrm{L}_\mathrm{dL} - \alpha^\mathrm{L}_\mathrm{dR}) \mathrm{e} V_\mathrm{L} = \SI{0.024}{} \times \mathrm{e} V_\mathrm{L} $ with the lever arms determined by finite bias measurements of the DQD.
The effective charge qubit coupling strength is $g^* = g \frac{2t_\mathrm{c}}{E_\mathrm{q}(\delta)}$ using the bare resonator coupling strength 
$$g/2\pi = \frac{\sqrt{\pi}}{2} \beta f_\mathrm{r} \sqrt{\frac{Z_\mathrm{r}}{h/e^2}} .$$
It is proportional to the lever arm difference of the coupling gate $\beta = \alpha^\mathrm{R}_{\mathrm{dR}} - \alpha^\mathrm{R}_{\mathrm{dL}}$, the resonator frequency $f_\mathrm{r}$, and the square root of the resonator impedance $Z_\mathrm{r}$.

\section{Fitting procedure}
\label{app:fit}

The parameters $g$, $\gamma$ and $2t_\mathrm{c}$ are determined by a least-squares fitting routine. 
We calculate the complex transmission $S_{21}(f_\mathrm{p}, \delta)$ from input-output theory for a given set of fit parameters $\theta = [g, \gamma, 2t_\mathrm{c}]$ and compute the mean squared error $Q(\theta)$ between calculation and measurement data.
The parameters given in the main text for each data set are the parameters $\hat{\theta}$ that minimize the error, found by sampling all three parameters over a wide range of values.
We numerically calculate the hessian matrix $H_{i,j} = \left. \frac{N}{Q(\hat{\theta})} \frac{\partial^2 Q(\theta)}{\partial \theta_i \partial \theta_j} \right|_{\hat{\theta}}$ around $\hat{\theta}$ with $N$ the size of $S_{21}$.
From $H$ we calculate the variance-covariance matrix $K = H^{-1}$ which contains information about uncertainties and correlations of our parameter estimates. 
%

%
For the dispersive case with $2 t_c/h \gg f_r$ the parameters that minimize the mean squared error are $g/2 \pi = 49.72 \pm \SI{0.27}{\MHz}$, $\gamma/2 \pi = 643 \pm \SI{24}{\MHz}$ and $2 t_\mathrm{c}/ h = 10.193 \pm \SI{0.046}{\GHz}$.
The strongest correlation is observed between $g$ and $2t_\mathrm{c}$ as expected from the definition of the effective coupling strength $g^*$ above, while the estimate of $\gamma$ is mostly uncorrelated to the other two parameters.
The relative uncertainty in our estimate of $\gamma$ is significantly larger compared to the uncertainty in the other two parameters.
This is because the resonator-qubit detuning is large over the whole range of $\delta$.
The incoherent dispersive interaction dominates the resonator response and does not depend strongly on the decoherence rate $\gamma$, rendering the mean squared error insensitive to the estimate of the decoherence rate. 
For the resonant case with $2 t_c/h < f_r$ the parameters that minimize the mean squared error are $g/2 \pi = 37.49 \pm \SI{0.19}{\MHz}$, $\gamma/2 \pi = 1.112 \pm \SI{0.008}{\GHz}$ and $2 t_\mathrm{c}/ h = 2.986 \pm \SI{0.015}{\GHz}$.
We observe stronger correlations between each pair of parameters compared to the dispersive case.
Furthermore, the uncertainty in $\gamma$ is smaller, because the resonant interaction is limited by charge qubit decoherence.
Therefore, the mean squared error is more sensitive to the estimate of the decoherence rate.

\bibliography{mybibliography}

\end{document}